\documentclass{aip-cp}
\pdfoutput=1
\usepackage[numbers]{natbib}
\usepackage{rotating}
\usepackage{graphicx}
\usepackage{lineno}

\begin{document}

\newcommand{\nonubb}  {$0 \nu \beta \beta$}
\newcommand{\twonubb}{$2 \nu \beta \beta$}
\newcommand{\bb}{$\beta\beta$}
\newcommand{\onecpRty} {\hbox{1~count/ROI/t-y}}
\newcommand{\MJ}{{\sc{Majo\-ra\-na}}}
\newcommand{\Dem}{{\sc{Demonstrator}}}
\newcommand{\MJD}{{\sc{Majorana Demonstrator}}}
\def\ppc{P-PC}  
\newcommand{\gam}{$\gamma$}
\def\nuc#1#2{${}^{#1}$#2}
\def\cpRty{c/(ROI t yr)}

\title{Low Background Materials and Fabrication Techniques for Cables and Connectors in the \MJD }

\author[duke,tunl]{M.~Busch\corref{cor1}}
\author[lbnl]{N.~Abgrall}
\author[uw]{S.I.~Alvis}
\author[pnnl]{I.J.~Arnquist}
\author[usc,ornl]{F.T.~Avignone~III}
\author[ITEP]{A.S.~Barabash}
\author[usd]{C.J.~Barton}
\author[ornl]{F.E.~Bertrand}
\author[mpi]{T.~Bode}
\author[lbnl]{A.W.~Bradley}
\author[JINR]{V.~Brudanin}
\author[uw]{M.~Buuck}
\author[unc,tunl]{T.S.~Caldwell}
\author[lbnl]{Y-D.~Chan}
\author[sdsmt]{C.D.~Christofferson}
\author[lanl]{P.-H.~Chu}
\author[uw,clara]{C. Cuesta}
\author[uw]{J.A.~Detwiler}
\author[sdsmt]{C. Dunagan}
\author[ut,ornl]{Yu.~Efremenko}
\author[ou]{H.~Ejiri}
\author[lanl]{S.R.~Elliott}
\author[unc,tunl]{T.~Gilliss}
\author[princeton]{G.K.~Giovanetti}
\author[ncsu,tunl,ornl]{M.P.~Green}
\author[uw]{J. Gruszko}
\author[uw]{I.S.~Guinn}
\author[usc]{V.E.~Guiseppe}
\author[unc,tunl]{C.R.~Haufe}
\author[lbnl]{L.~Hehn}
\author[unc,tunl]{R.~Henning}
\author[pnnl]{E.W.~Hoppe}
\author[unc,tunl]{M.A.~Howe}
\author[blhill]{K.J.~Keeter}
\author[ttu]{M.F.~Kidd}
\author[ITEP]{S.I.~Konovalov}
\author[pnnl]{R.T.~Kouzes}
\author[ut]{A.M.~Lopez}
\author[queens]{R.D.~Martin}
\author[lanl]{R. Massarczyk}
\author[unc,tunl]{S.J.~Meijer}
\author[mpi,tum]{S.~Mertens}
\author[lbnl]{J.~Myslik}
\author[unc,tunl]{C. O'Shaughnessy}
\author[unc,tunl]{G.~Othman}
\author[lbnl]{A.W.P.~Poon}
\author[ornl]{D.C.~Radford}
\author[unc,tunl]{J.~Rager}
\author[unc,tunl]{A.L.~Reine}
\author[lanl]{K.~Rielage}
\author[uw]{R.G.H.~Robertson}
\author[uw]{N.W.~Rouf}
\author[unc,tunl]{B.~Shanks}
\author[JINR]{M.~Shirchenko}
\author[sdsmt]{A.M.~Suriano}
\author[usc]{D.~Tedeschi}
\author[unc,tunl]{J.E.~Trimble}
\author[ornl]{R.L.~Varner}
\author[JINR]{S. Vasilyev}
\author[lbnl]{K.~Vetter}
\author[unc,tunl]{K.~Vorren}
\author[lanl]{B.R.~White}
\author[unc,tunl,ornl]{J.F.~Wilkerson}
\author[usc]{C. Wiseman}
\author[usd]{W.~Xu}
\author[JINR]{E.~Yakushev}
\author[ornl]{C.-H.~Yu}
\author[ITEP]{V.~Yumatov}
\author[JINR]{I.~Zhitnikov}
\author[lanl]{B.X.~Zhu}

\affil[duke]{Department of Physics, Duke University, Durham, NC, USA}
\affil[tunl]{Triangle Universities Nuclear Laboratory, Durham, NC, USA}
\affil[lbnl]{Nuclear Science Division, Lawrence Berkeley National Laboratory, Berkeley, CA, USA}
\affil[uw]{Center for Experimental Nuclear Physics and Astrophysics, and Department of Physics, University of Washington, Seattle, WA, USA}
\affil[pnnl]{Pacific Northwest National Laboratory, Richland, WA, USA}
\affil[usc]{Department of Physics and Astronomy, University of South Carolina, Columbia, SC, USA}
\affil[ornl]{Oak Ridge National Laboratory, Oak Ridge, TN, USA}
\affil[ITEP]{National Research Center ``Kurchatov Institute'' Institute for Theoretical and Experimental Physics, Moscow, Russia}
\affil[usd]{Department of Physics, University of South Dakota, Vermillion, SD, USA} 
\affil[mpi]{Max-Planck-Institut f\"{u}r Physik, M\"{u}nchen, Germany}
\affil[JINR]{Joint Institute for Nuclear Research, Dubna, Russia}
\affil[unc]{Department of Physics and Astronomy, University of North Carolina, Chapel Hill, NC, USA}
\affil[sdsmt]{South Dakota School of Mines and Technology, Rapid City, SD, USA}
\affil[lanl]{Los Alamos National Laboratory, Los Alamos, NM, USA}
\affil[ut]{Department of Physics and Astronomy, University of Tennessee, Knoxville, TN, USA}
\affil[ou]{Research Center for Nuclear Physics, Osaka University, Ibaraki, Osaka, Japan}
\affil[princeton]{Department of Physics, Princeton University, Princeton, NJ, USA}
\affil[ncsu]{Department of Physics, North Carolina State University, Raleigh, NC, USA}	
\affil[blhill]{Department of Physics, Black Hills State University, Spearfish, SD, USA}
\affil[ttu]{Tennessee Tech University, Cookeville, TN, USA}
\affil[queens]{Department of Physics, Engineering Physics and Astronomy, Queen's University, Kingston, ON, Canada} 
\affil[tum]{Physik Department, Technische Universit\"{a}t, M\"{u}nchen, Germany}
\affil[clara]{Present Address: Centro de Investigaciones Energ\'{e}ticas, Medioambientales y Tecnol\'{o}gicas, CIEMAT, 28040, Madrid, Spain}
\corresp[cor1]{Corresponding author: matthew.busch@duke.edu}

\maketitle


\begin{abstract}
The \MJ\ Collaboration is searching for the neutrinoless double-beta decay of the nucleus $^{76}$Ge. The \MJD\ is an array of germanium detectors deployed with the aim of implementing background reduction techniques suitable for a tonne scale $^{76}$Ge-based search (the LEGEND collaboration). In the \Dem, germanium detectors operate in an ultra-pure vacuum cryostat at 80 K. One special challenge of an ultra-pure environment is to develop reliable cables, connectors, and electronics that do not significantly contribute to the radioactive background of the experiment. This paper highlights the experimental requirements and how these requirements were met for the \MJD, including plans to upgrade the wiring for higher reliability in the summer of 2018. Also described are requirements for LEGEND R\&D efforts underway to meet these additional requirements
\end{abstract}

\section{INTRODUCTION}

The \MJD\ \cite{abg14} is an experiment currently taking data in the Davis Campus of the Sanford Underground Research Facility (SURF) located at the 4850-foot level of the former Homestake Mine in Lead, SD, USA. The goal of the experiment is to demonstrate backgrounds low enough to justify building a tonne scale experiment, to establish the feasibility to construct and field modular arrays of Ge detectors, and to search for additional physics beyond the standard model. The \MJ\ Detector Room at the David Campus houses the experiment in class 1000 or better cleanroom conditions. The experiment consists of 2 modular arrays of high purity germanium detectors in identical high purity vacuum cryostats. There are 29 detectors in each cryostat, resulting in a total installed detector mass of 44.1 kg, in which 29.7 kg of these detectors are produced from material that is 88\% enriched in $^{76}$Ge. The remaining 14.4 kg of detectors are made from Ge with natural isotopic abundance ratios. The detectors in each cryostat are arranged in a close packed array of 7 columns called strings. Each string is a vertical stack of 3-5 detectors depending on detector height. The 2 modules are housed in a graded compact shield consisting of underground electroformed copper, specially sourced commercial copper \cite{abg16a}, specially prepared lead bricks \cite{abg16a}, a sealed aluminum enclosure to isolate room air, scintillating acrylic muon veto panels, and a polyethylene shield. Figure \ref{fig:1}(a) shows a cross section schematic of the experiment cutting though the cryostats to provide a view of the internal detectors, cryostat, and graded shield.

Each cryostat is made primarily of ultra clean underground electroformed copper and specially screened insulating materials such as NXT-85 fluoropolymer and Vespel SP-1 polyimid. The collaboration has achieved very stringent contamination level goals in all components \cite{abg16a}. This paper describes the special challenges of achieving these goals for the in-vacuum cables and connectors of the experiment.

\section{CABLE AND CONNECTORS}

\subsection{Configuration and Functional Requirements}

\begin{figure}[t]
\centering
\begin{tabular}[b]{c}
  \includegraphics[width=0.45\textwidth]{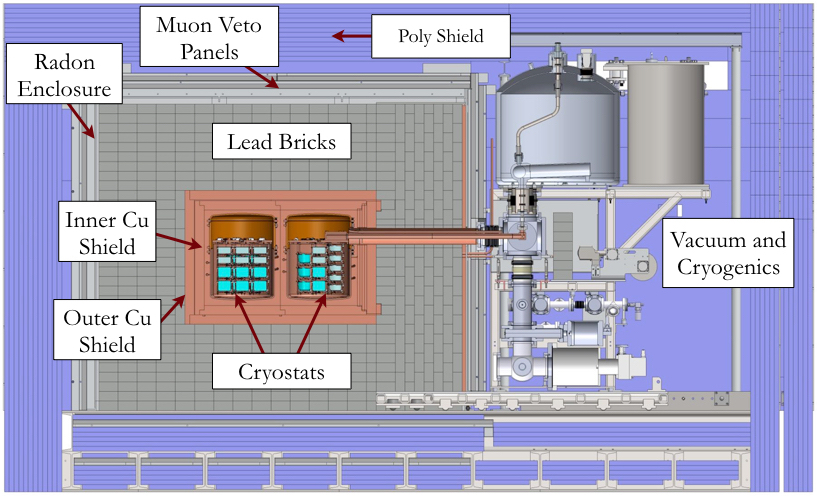} \\
  \small (a)
\end{tabular}
\begin{tabular}[b]{c}
  \includegraphics[width=0.5\textwidth]{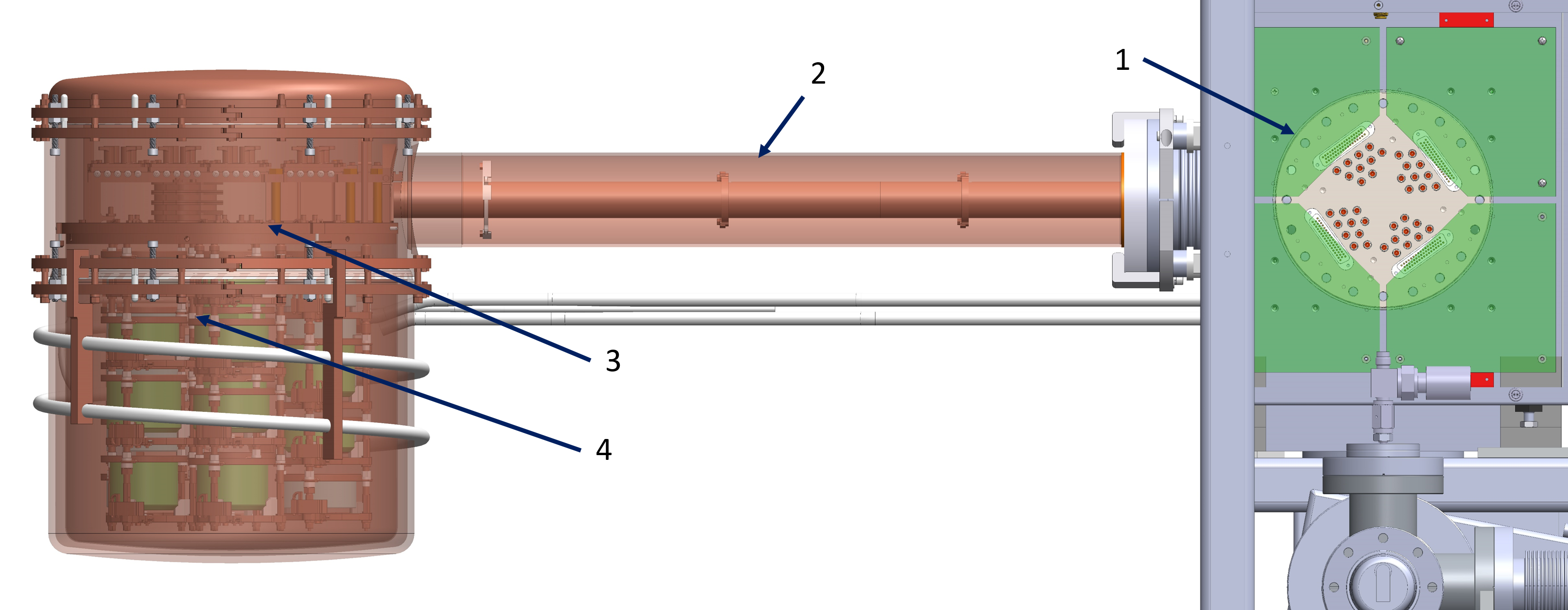}\\
  \small (b) 
  \end{tabular}
  \caption{(a) Section view schematic of the \MJD\ experiment. (b) Module only view showing 1. Vacuum feedthrough cube, 2. Cross arm, 3. Cold plate, 4. Cold plate to detector cables.}
  \label{fig:1}
\end{figure}

In-vacuum cables for the \Dem\ run from the vacuum feedthrough cube, through the cross arm and onto the cold plate of the cryostat (see Fig. \ref{fig:1}(b)). From the cold plate there is a dedicated pass through hole for each string. Any excess cable slack is arranged on top of the cold plate. The total cable length for each cable including necessary slack for assembly is 2.15 m.

Each detector requires 4 individual signal cables with independent ground shields. The signal cable must have at least 50 $\Omega$ impedance and low capacitance to be compatible with the front end electronics mounted at the detector. Each detector also needs an individual shielded high voltage cable capable of holding 5 kVDC at a few microamps max current. Typical operating conditions are 10-100 pA at 3-4.5 kVDC. Each ground shield is individually grounded in the electronics box and is isolated from the rest of the cryostat. 

Cable harnesses can be fastened to cold surfaces along the cross arm and then inserted as an assembly. To minimize the diameter of the cross arm for effective shielding and to provide a practical connector size, threading cables through the cross arm at detector installation was deemed impractical. As a result, there must be high purity low mass connectors at the cold plate or at the detector for each cable.

\subsection{Radiopurity Requirements for Cables and Connectors}

The most challenging requirements of the signal and high voltage cables are that they have very low radioactivity. Several commercial, modified commercial, and handmade options were investigated during development. All commercial options failed our assay requirements \cite{abg16a}. Handmade options passed assay requirements, but could not be made to pass electrical performance requirements. A customized commercial solution was found to meet all requirements. Axon' Cable \cite{axon} produces medical and space grade cable, and was able to customize standard cables to meet our radiopurity requirements. Axon' pico-coax is a very fine gauge coaxial cable that can be made with high purity bare OFHC C10100 copper wire and unpigmented FEP insulation. The raw materials met our background goals and a test of the finished product also met our goals with some minor changes to Axon' medical grade cleanliness protocols. Performance requirements for these custom cables were verified at each stage of production \cite{abg16b}. Key specifications of our Axon' cables are listed in Table \ref{tab:1}. Specific activities for Axon' signal cable are listed as upper limits using component material assay limits \cite{abg16a}.

\begin{table}[h]
\caption{Key specifications for Axon' cables}
\label{tab:1}
\tabcolsep7pt\begin{tabular}{lll}
\hline
Parameter 	& Axon' custom 	& Axon' custom \\
			& HV cable		& signal cable \\
\hline
Central conductor dia. (mm) &	0.152 & 	0.076\\
Jacket OD (mm) &			1.2 &	 	0.4	\\
Specific mass (g/m) &		3.0 & 	0.4 	\\
Max voltage (kVDC) & 		5.0 & 	Not rated \\
Max Current ($\mu$A) &		20 & 		Not rated \\
Capacitance (pF/m) &		Not rated 	& 	87  \\
Impedence ($\Omega$) &		Not rated & 50  \\
Specific activity $^{232}$Th ($\mu$Bq/m) & 	0.007 & 	$<$0.03 \\
Specific activity $^{238}$U ($\mu$Bq/m) & 	0.43 & 	$<$0.13 \\
\hline
\end{tabular}
\end{table}

\subsection{Detailed Design of Cables and Connectors}

In-vacuum connectors near the detectors are limited by the materials in approved for use \cite{abg16a} but also by the need to terminate our customized Axon' cables. No commercial connectors are appropriate for use with this cable. Axon' typically solders wire to a circuit board and uses board mount connectors for other customer applications. Early prototypes in development used this principle in an attempt to make signal and high voltage connectors, but were not successful due to HV breakdowns, ergonomic challenges, and poor performance at low temperature. For the high voltage cables \cite{abg16b}, a forked style spade lug is clamped onto the detector HV ring. This ``HV fork" serves as a demountable connector directly at the detector. The fork itself is produced underground from our underground electroformed copper, and the termination is made with a diminutive 3 mg clamp pin made of Vespel SP-1. Two additional holes are drilled in the HV fork to serve as a strain relief. The appropriate strip back length has been experimentally determined \cite{abg16b} at 1.5 - 2 cm.
At the feedthrough side of the HV cable, the radiopurity requirements are greatly relaxed, as the feedthrough flange is outside of the lead shield. However, a custom connector is still required \cite{abg16b}, as standard connectors are not available for our custom Axon' cable. A 2-pin PEEK body is custom machined to house commercial Be-Cu D-sub sockets. The terminations are soldered or silver epoxied into the sockets. Having 2 identical pins, one for ground and one for HV allows for a demountable connector and the ability to individually ground each HV ground braid, which became vital for commissioning.

Signal connectors require more insulation mass to isolate the 4 central wires and 4 independent ground shields. Hence this connection is made at the top of the cold plate, where there is some shielding effect from the cold plate and from physical distance to the detectors. There is also more room to make and organize connections. Vespel SP-1 housings were machined in cleanroom conditions, and special-run Mill-Max \cite{mm} commercial gold plated brass pins were produced without Be-Cu spring inserts. Contact spring force is provided by offsetting the male and female connections radially, hence bending the male pins in place and using the spring constant of the pin itself to maintain contact. Terminations are hand-soldered under a microscope in cleanroom conditions. In practice, the tolerances required for reliable contact through the required temperature range was very difficult to achieve, and failed in the field on several occasions. The large investment in time and materials involved in producing and soldering these connectors limited our ability to reject flaky connectors during testing. Testing through temperature cycling was also difficult to reproduce without installation in the cryostat.

As is the case with the HV connections, the signal cable feedthrough side has relaxed radiopurity requirements but must still be customized to accommodate Axon' cables. Commercial vacuum compatible PEEK 50 pin D-Subs were modified by cutting slots between adjacent pins. This allowed for a single cable to be laced through the connector back shell for independent grounding of each ground shield to a D-sub socket. Some of these connections failed in the field after successful testing on the bench. This could be due to rough handling at installation and/or poor strain relief, or due to deflection of the connector body itself, preventing positive contact in the central region of the connector.

\section{FUNCTIONAL PERFORMANCE OF INSTALLED CABLES AND CONNECTORS}

The design and fabrication of cables and connectors met all requirements, but at installation some functional failures limited the success of these designs. As a result of these failures, several installed detectors are not collecting data in the experiment to date.

\begin{table}[h]
\caption{Summary of cable and connector performance in the \MJD}
\label{tab:2}
\tabcolsep7pt\begin{tabular}{lll}
\hline
\# of detectors &			Operating condition & 			Root cause \\
\hline
58 & 					installed in \Dem\ cryostats &		\\
29 &					operating normally & 				\\
12 &					operating with HV ground isolated & 	HV cable handling / robustness \\
9 & 					Unable to bias & 				Signal connection failure \\
7 & 					Unable to bias & 				HV connection or damage \\
1 &					Unable to bias due to high leakage current & 	Detector damage at assembly \\
\hline
\end{tabular}
\end{table}

Table \ref{tab:2} summarizes the status of installed detectors in the \Dem. The 12 detectors operating with isolated ground shields exhibited discharge events and their bias was lowered until discharges stopped. It should be noted that every cable was tested for 24+ hours in a test vacuum chamber after termination as part of our quality control process \cite{abg16b}. After some experimenting, the ground shields were disconnected from ground at the electronics box and re-biasing showed good performance with no discharges and no increased noise in the region of interest.

Further testing on the bench has shown that HV cables that are pinched exhibit this same behavior of recovering HV standoff with the ground shield isolated from ground. It was also shown in testing that a sharp edge and a small amount of pressure can easily lead to permanent damage that reduces the maximum voltage before breakdowns occur.

One likely cause of HV cable damage at installation is at the baffle plates. These baffle plates provide physical shielding between the inner copper shield and feedthrough cube while maintaining a reasonable pumping speed between the vacuum pumps below the feedthrough cube and the cryostat. The inside edges are fairly sharp, and any handling during installation can cause a cable to be pinched at the edge of this feature.

\section{CABLE AND CONNECTOR UPGRADE}
U.S. Department of Energy (DOE) funds have been provided to upgrade the internal cables and connectors for the \MJD. The goal of the upgrade is to demonstrate that at least 90\% of all installed detectors can be operated in an ultra-low background environment. The following section highlights progress and plans for this upgrade.

\subsection{HV Cable Protection }

Bundling cables together and providing rounded edges at all interface points will greatly reduce the likelihood of cable damage at installation. Having a large opening for a cable bundle provides a line of sight for $\gamma$ radiation, so a new approach to baffle plate arrangement is required. A helical cable routing path provides reasonable pumping speed while eliminating line of site paths. The Prototype Cryostat, made of commercial copper but with identical dimensions to the \Dem\ cryostats, has been installed at the University of North Carolina (UNC). Test plates have been manufactured and installed in the Prototype Cryostat to confirm that the pumping speed is acceptable. These test plates will be used to practice and verify that the installation methods do not damage the HV or signal cable harnesses at assembly.

Continuous lacing is a standard method for organizing and reducing the risk of damage during assembly. Properly wrapped cable assemblies also provide some strain relief at termination points. NASA \cite{nasa} has developed a standard for proper cable wrapping. Using this guide, a promising wire wrap material has been demonstrated to be effective. Zeus \cite{zeus}  ePTFE 0.0124" diameter suture monofilament is intended for human surgical stitches. It is easy to tie and has high strength and good elasticity properties. It is made from a material that has a history of acceptable assay results, and is produced and packed in medical grade cleanroom conditions. An assay campaign is planned to verify the suitability of this material.

\subsection{HV Cable Connectors}

Three detectors installed in Module 2 have no HV response. The most likely cause of this failure is a low quality connection of the HV wire at the HV fork, which depends on an interference fit between a drilled hole and the Vespel clamp pin. Several of these pins were found to be loose at installation for Module 2 and repaired on site by replacing the pins. The diameter of the pin varies slightly from sample to sample and a loose fit will be exacerbated at low temperatures due to differential rates of thermal contraction. To remedy this possible failure scenario, a modified HV fork design has been produced and tested using a direct crimp method. Initial testing looks good, and a full test with an operating detector and an assay campaign are planned.

The feedthrough connector for the HV cables is not suspected as a cause for inoperable detectors, but new custom connectors must be produced for the rewire. This has provided an opportunity to improve the handling ergonomics and robustness of this connector as a precaution. A different commercial socket provided increased connector retention force on the feedthrough pins, reducing the risk of accidental unplugging during installation. A larger PEEK housing prevents the possibility of shorting HV to ground with stray wires. A redesigned strain relief reduces the risk of pinching the cable at the strain relief, which is a possible secondary cause of HV cable damage.

\section{SIGNAL CONNECTORS}
Poor connection quality in the signal connectors is a major cause of inoperable installed detectors. The fabrication and termination of these connectors was also a large investment of technician labor, so new designs have been developed to provide higher reliability and lower labor investment while maintaining or improving on the excellent assay results of the currently installed solution.

Custom Interconnects \cite{ci} has collaborated on a revised design of the signal connectors that incorporate gold plated molybdenum fuzz buttons. A fuzz button looks like microscopic steel wool and behaves like a springy conductive sponge at the full range of operating temperatures required for use in the \Dem. These fuzz buttons have extremely low mass of ~1 mg each, and show promise as a low background material. An assay campaign is planned to verify this. A phosphor bronze retaining clip has been added to the design to provide secure engagement, since pins are no longer offset to provide engagement. This phosphor bronze is from the same stock used and assayed \cite{abg16a} for the spring washers to retain the HV forks at installation. The fuzz button design has performed well in initial LN testing, and a test is planned with operating detectors in the Prototype Cryostat.

Existing signal connectors were hand soldered under a microscope by a single technician as a way to provide consistent quality, solder mass, and flux removal. This method has been deemed impractical for a rewire effort based on technician availability. An alternate method has been investigated and shows promise in terms of connection quality and repeatability. A Sunstone \cite{sw} Orion 150s micro-TIG and resistance welder has been purchased by UNC for development purposes. This method also reduces the total mass and potentially the background budget for the connectors by completely eliminating the use of solder and flux, though the connectors made in this fashion will need to be assayed to confirm the lower backgrounds.

Feedthrough side connectors for signal cables are not suspected as a root cause for inoperable detectors. However, several failures were observed at installation and these failures significantly reduced the number of available spare circuits installed in the cross arm. Installed feedthrough connectors use solder or silver epoxy to terminate signal cables into a modified commercial D-sub. Crimped D-sub sockets are recommended for MIL-Spec and space grade connections over soldered or silver epoxy \cite{nasa}. High quality 8-lobe Mil-Spec crimping tools are a key to producing repeatable crimped connections. Crimped sockets have been successfully tested using an Astro-Tool \cite{atc} crimper with our pico-coax signal cables; no such crimping option could be found when the connectors were first developed. Using crimped sockets allows for the use of a repairable space grade D-Sub connector from Glenair \cite{gle}. Such a connector will undergo a full test in the Prototype Cryostat.

\section{LEGEND CABLE AND CONNECTOR DEVELOPMENT}

The LEGEND experiment is planned as a next generation germanium-based neutrinoless double beta decay experiment. R\&D funds have been awarded by the DOE to develop cable and connector technology required to continue reducing backgrounds in Ge based experiments. The LEGEND experiment will combine the expertise and best design elements of the \MJD\ \cite{abg14} and GERDA \cite{ago17} experiments. The electronics readout scheme is not finalized for this experiment, so cable and connector functional requirements are not yet established. The default plan for LEGEND is to configure the experiment similar to GERDA using materials and processes developed by the \MJD\ to reduce backgrounds to well below what was achieved in either experiment.
One feature of the GERDA cables and connectors that will most likely be adopted to LEGEND is the use of wire bonded contacts directly on the surface of the germanium detectors. 
LEGEND cable R\&D work will focus on developing a new HV cable with higher durability and a higher safety factor for maximum voltage using cleaner starting copper wire. The \Dem\ rewire signal connectors and HV fork designs will be modified to work within the LEGEND configuration, and additional signal connector and feedthrough connector options will be investigated and assayed to develop suitable options.

\section{CONCLUSIONS}

Sacrifices in connection quality, termination reliability, and cable durability made in order to satisfy radio-purity requirements are leading causes of inoperable detectors in the \MJD. Development work is underway to upgrade the terminations and improve cable harness ruggedness in the \Dem. This connector development work feeds directly into design and development work for the LEGEND experiment, as well as benefitting the low background community at large.

\section{ACKNOWLEDGMENTS}
The author would like to thank the superCDMS collaboration for the suggestion of using fuzz buttons in a low background low temperature environment.
This material is based upon work supported by the U.S. Department of Energy, Office of Science, Office of Nuclear Physics, the Particle Astrophysics and Nuclear Physics Programs of the National Science Foundation, and the Sanford Underground Research Facility. 

\bibliographystyle{aipnum-cp}%
\bibliography{/Users/guiseppe/work/latex/bib/mymj}%

\end{document}